\documentstyle[11pt,newpasp,twoside]{article}
\def\edcomment#1{\iffalse\marginpar{\raggedright\sl#1\/}\else\relax\fi}
\reversemarginpar

\def\Msun{\hbox{{\it M$_\odot$}}}
\def\Minit{\hbox{M$_{\rm initial}$}}

\def\simgr{\mathrel{\hbox{\rlap{\hbox{\lower4pt\hbox{$\sim$}}}\hbox{$>$}}}}
\def\simls{\mathrel{\hbox{\rlap{\hbox{\lower4pt\hbox{$\sim$}}}\hbox{$<$}}}}

\def\arcsec{\hbox{$^{\prime\prime}$}}

\begin{document}
\title{Starburst Clusters in Galactic Nuclei}
 \author{Donald F. Figer}
\affil{STScI, 3700 San Martin Drive, Baltimore, MD 21218; \\JHU, 3400 Charles 
Street, Baltimore, MD 21218}
\author{Mark Morris}
\affil{UCLA, Division of Astronomy, 405 Hilgard Avenue, LA CA 90095-1562}

\begin{abstract}
Galactic nuclei often harbor a disproportionately large amount 
of star formation activity with respect to their surrounding disks. 
Not coincidentally, the density of molecular material in galactic
nuclei is often also much greater than that in disks (Table 1 in Kennicutt 1998). 
The interplay between rich populations of young stars and dense molecular 
environments is evident in our own Galactic center, which hosts over 10\% of 
Galactic star formation activity within only $<$0.1\% of the volume of the 
Galactic disk. Data obtained
with the VLA and HST reveal a variety of star forming sites in the Galactic
Center, including a substantial population of stars that are formed in very
dense and massive clusters, while other stars are formed in somewhat sparsely 
populated associations of massive stars. Indeed, three of the stellar clusters 
are the most massive and densest in the Galaxy. 
In this paper, we discuss the Galactic center environment and its compact 
young star clusters, and compare them to their counterparts in star forming 
galactic nuclei, concluding that dense molecular environments and large 
velocity dispersions combine to alter star formation activity in both cases, 
particularly as regards massive young clusters.
\end{abstract}

\section{Introduction}
While occupying just a tiny fraction of the Galactic disk, the central molecular 
zone (CMZ, r$_{\rm Gal}$ $<$ 500~{\it pc})
of our Galaxy harbors a plethora of astrophysical phenomena. Not only is 
about 10\% 
of the star formation activity of the Galaxy inferred to be taking place there, but 
$\sim$10\% of the
Galaxy's molecular gas is present there, indicating that while the star formation 
rate per unit mass of gas is the same as in the disk (G\"usten 1989), the star 
formation rate
per unit volume is orders of magnitude higher than in the disk. The 
form that the star formation takes can apparently be dramatically different than 
in the disk. 
While the majority of newborn stars in the CMZ were perhaps formed in typical 
clusters or associations like
those found in the disk, most of the young stars ($\tau_{\rm age} < 10^7 yr$) 
currently seen in 
the core of the CMZ (r$_{\rm Gal}$ $<$ 50) were
formed in three spectacular, dense clusters, the most massive such clusters in 
the Galaxy. 
Recently, the Hubble Space Telescope has 
identified similar, and even more massive, young clusters in other galaxies. 
While not always
confined to galactic nuclei, they are often found there. Evidently,
such clusters represent the results of an important mode of star formation, for 
our Galaxy and others, and we hereafter refer to them as starburst clusters. In 
this paper, we describe recent results concerning the starburst cluster 
population
in the Galactic center and discuss how these clusters compare to similar 
clusters
in other Galactic nuclei. 

\section{Star Formation in the Galactic center -- Theory}
The gravitational collapse of a clump of material within a molecular cloud to 
form a 
star is slowed, or even counteracted, by a number of factors, including thermal 
gas 
pressure gradients, a magnetic field which permeates the molecular material, 
turbulence or other forms of macroscopic internal velocity dispersion, and the 
outward pressure of radiation from the heated clump. 
Most of these factors appear to be particularly important in the Galactic center, 
and we expect them to have a pronounced effect on both the rate of star 
formation there and on the initial mass function (IMF). 

Indeed, Morris (1993) notes that the Jeans mass, $M_J$, in the
GC is extraordinarily large ($\sim$10$^5$~\Msun), whether calculated by 
invoking
turbulent velocities ($M_J~\propto~\Delta V_{\rm turb}^{3}~n^{-1/2}$) or 
magnetic fields
($M_J~\propto~B^3 n^{-2}$) as the means of support against collapse, where n 
is the 
particle density in the cloud. (Indeed, the kinetic and magnetic energies within 
clouds appear to be close to equipartition.) The physical meaning of such a 
large Jeans mass is unclear, except that it suggests a
tendency for the medium to favor the formation of relatively massive stars, 
compared
to clouds in the Galactic disk, which have a much smaller energy density. 
Furthermore,
it is perhaps inappropriate to relate observed macroscopic quantities (turbulent 
velocities or magnetic field geometries) to processes such as compression 
and collapse, which are likely to be operating on smaller, unresolved scales. 

A similar problem is found in the early Universe during the time of 
recombination, wherein the Jeans mass is also $\sim$10$^5~\Msun$. In this 
case, Peebles \& Dicke (1968) note that this is about the mass of a typical globular cluster, so they suggest that the masses of globular clusters are defined by the original Jeans instability criterion.  Subsequent compression,
fragmentation, and cooling significantly alter the quantities which enter into
the Jeans mass equation, giving ultimately the spectrum of stellar masses which
form within the clusters. 
Perhaps a similar argument applies to the massive star clusters
in the Galactic center, given that their masses (a few 10$^4~\Msun$) are not
too different than the Jeans mass in the GC. 

\section{Starburst Clusters in the Galactic center -- Observations}
A number of infrared studies in the past 30 years have revealed three starburst
clusters in the Galactic center: 1) the Central cluster (Becklin \& Neugebauer 
1968; Forrest et al.\ 1987; Krabbe et al.\ 1991;
Allen 1994; Libonate et al.\ 1995; Tamblyn et al.\ 1996; Najarro et al.\ 1997; Genzel et al.\ 2000; 
Paumard et al.\ 2000), 2) the Quintuplet cluster (Okuda et al.\ 1990; Nagata et 
al.\ 1990; Glass, Moneti, \& 
Moorwood 1990; Figer, McLean, \& Morris 1999a), and 
3) the Arches cluster (Nagata et al.\ 1995; Cotera et al.\ 1996; 
Serabyn, Shupe, \& Figer 1997; Figer et al.\ 1999b). 
Indeed, radio observations obtained over the same time period hinted at these 
clusters by revealing 
their associated ionizing radiation in the form of HII regions, i.e. the ``mini-
spiral,''
the ``Sickle'', and the ``Thermal Arched Filaments.''
Each cluster is thought to contain a few thousand stars and to have masses of 
at least 10$^4$~\Msun. The Arches
and Quintuplet clusters have been probed down to a few solar masses (Figer et 
al.\ 1999b), while data for the
Central cluster are only available for stars at the very upper tip of the main 
sequence 
(Genzel et al.\ 1997; Eckart, Ott, \& Genzel 1999; Figer et al.\ 2000). 
The clusters are rather compact, with half-light radii between 1~{\it pc} (the 
Quintuplet) and 0.2~{\it pc} (the Arches cluster; Figer et al.\ 1999b). Note that this 
implies a central density 
of $>5(10^5)~\Msun~{\it pc}^{-3}$ for the Arches cluster, greater than that for most 
old globulars.

\begin{figure}
\vskip 0.75in
\plotfiddle{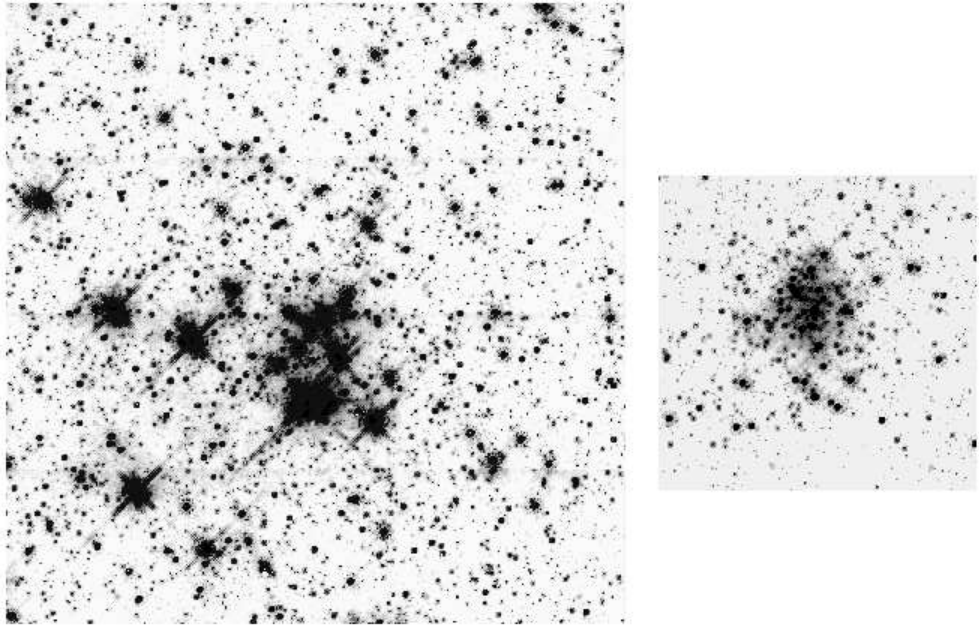}{2.5in}{-90}{50}{50}{-200pt}{260pt}
\vskip .2in
\caption{HST/NICMOS images of the Quintuplet ({\it left}) and Arches ({\it right}) 
clusters (Figer et al.\ 1999b). The images are plotted to the same scale, where the
Arches cluster image spans $38\arcsec\times38\arcsec$, or 1.5~{\it pc} on a side at the distance
of the GC (8000~{\it pc}).}
\end{figure}
The presence of these extraordinary clusters makes it possible to measure the 
IMF via a direct count of coeval stars in the Galactic center. Recall that the IMF 
describes the relative number of stars produced in a star forming event
as a function of initial stellar mass, and is often expressed as a single power 
law
over mass ranges above 1~\Msun, with a form d(Log N)/d(Log \Minit) 
= ${\rm \Gamma}$ . The IMF for most young clusters can be described  
reasonably well by a power law with Salpeter index (= $-$1.35 ; Salpeter 1955), 
although significant variations
are observed ($-$0.7~$>$~${\rm \Gamma}$~$>$~$-$2.1) (Scalo 1998). 

Figer et al.\ (1999b) targetted the Arches and Quintuplet clusters for just such a 
measurement,
avoiding the prohibitively confused Central cluster. They describe 
{\it Hubble Space Telescope} (HST) Near-infrared Camera and Multi-object 
Spectrometer (NICMOS)
observations which were used to identify main sequence stars in the Galactic 
Center with initial masses well below 10~\Msun, leading to the first 
determination of the IMF for any population in the Galactic center. They found 
a slope which is significantly greater than $-$1.0 (see Figure 2), and so is one 
of the flattest mass functions ever observed for \Minit~$>$~10~\Msun, although 
note that Eisenhauer et al.\ (1998) found a similar result for the Galactic cluster 
NGC3603. These two results can be contrasted with the average IMF slope for 
30 clusters in the Milky Way and LMC: $\approx$$-$1.3 for log(\Minit)~$>$~1, 
although ${\rm \Gamma_{\rm NGC6611}}$=$-$0.7$\pm$0.2 
and ${\rm \Gamma_{\rm NGC2244}}$=$-$0.8$\pm$0.3 over this mass range 
(Scalo 1998).
Some of these clusters discussed in Scalo (1998) suggest a flattening of the 
IMF at higher masses, although the IMF slopes reported for these comparison 
clusters are in general biased toward lower masses.  
Finally, we find that there are $\simgr$10 stars with \Minit~$>$~120~\Msun\ in 
the Arches cluster. This number is consistent with the absence of any clear 
upper-mass cutoff to the IMF.

\begin{figure}
\vskip 0.75in
\plotfiddle{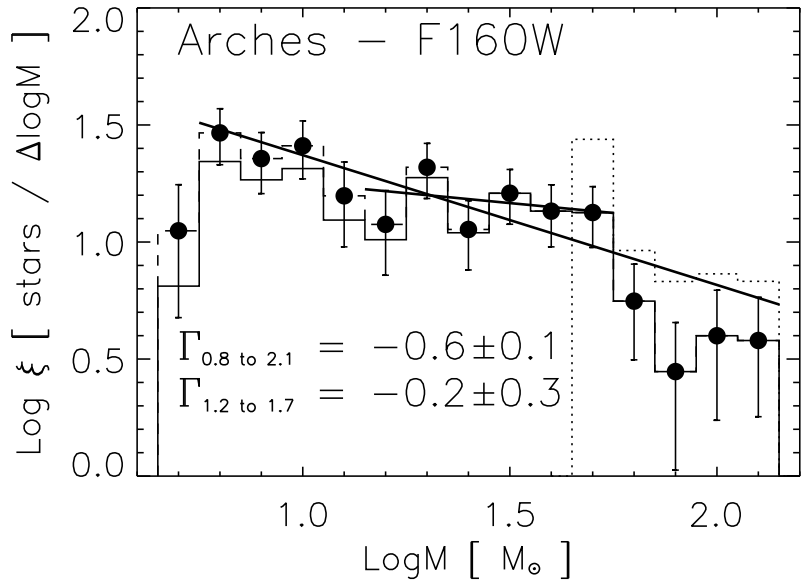}{2.0in}{0}{70}{70}{-220pt}{-280pt}
\vskip .2in
\caption{Mass function for Arches cluster as measured in the F160W image for stars within
an annulus from 3\arcsec\ to 9\arcsec\ (reproduced from Figer et al.\ 1999b). 
Lines have been fit to the completeness-corrected data ({\it dashed lines}) over two mass ranges. 
The slopes of both are 
relatively flat. The dotted histogram shows the number of massive stars over the 
whole cluster, including the inner region; the counts have been terminated at the mass
where incompleteness exceeds 50\%.}
\end{figure}

The rarity in our Galaxy of massive compact clusters, such as are found in the 
Galactic center, suggests that they form under circumstances which are quite 
unusual. Locally, their formation must have been under conditions which fall 
under the category of ``starburst,'' given the tremendous energy release 
accompanying the coeval formation of $\sim$10$^4$~\Msun\ of predominantly 
massive stars. Furthermore, although the central density of the Arches cluster 
may have been enhanced somewhat by dynamical processes since its 
formation, its equivalent density of 10$^7$ hydrogen molecules per cm$^3$ is 
as high as the density of only the densest of cloud cores, suggesting that the 
formation process must have been very efficient. One imagines the sudden 
and catastrophic transformation of an entire dense cloud into a massive, 
compact star cluster having an unusually flat IMF. While such clusters merit the 
name of ``starburst cluster,'' they would presumably appear as a starburst on a 
Galactic scale, and thus be an element of what we would regard as a starburst 
galaxy, only if a multitude ($\simgr$10 -- 100) of such clusters were to form all at 
once in a galactic nucleus.

\section{Starburst Clusters in Other Galactic Nuclei}
Populations of starburst clusters in galactic nuclei are seen in a variety of 
galaxies,
some of which display general starburst or AGN properties. These galaxies 
include 
NGC 1275 (Carlson et al.\ 1998), Arp 220 (Scoville et al.\ 1998), NGC 253 
(Watson et al.\ 1996), 
NGC 1365 (Lindblad 1999), NGC 2903 (Mulchaey \& Regan 2001), among 
many others. Of course, many galactic nuclei, such as M31, have little or no star 
formation activity and certainly no starburst clusters. In galaxies where clusters 
have been produced in large numbers, their 
properties are similar to those inferred for clusters in our own Galactic globular 
cluster system in its infancy, i.e., both types of systems might share 
commonality in their formation histories (Whitmore 2000). 

Recent evidence suggests that nuclear starbursts occur in the late stages of 
galactic interaction
events, a beautiful example of which can be seen in the ``Antennae'' merger 
system (Whitmore et al.\ 1999).
Indeed mergers or interactions produce starbursts in many cases, as is seen 
observationally 
(Borne et al.\ 2000) and anticipated theoretically (Mihos 1999). Mihos \& 
Hernquist (1996) performed
Smoothed Particle Hydrodynamics calculations and produced a simulation  
which reproduces many of the
dynamical features thought to describe a galaxy merger event such as the one 
leading to the ``Antennae''
system. In this simulation, star formation starts in the arms and finishes in the 
centers of the galaxies
after mass inflow. The predicted timeline produces a range of cluster ages and 
demonstrates the
importance of galaxy interactions in producing a spectrum of cluster masses; 
however, such dramatic events as mergers are not requisite for the production 
of
starburst clusters, at least on a small scale, as evidenced by our own Galactic 
center.

\section{Conclusions}
The starburst clusters in our Galactic center are similar to those found in the 
centers of
some starburst galaxies, although they are much fewer in number than those 
found in galaxies
culled for their starburst properties. Of course, this is simply a selection effect, 
in that 
galaxies with the highest star formation rates are most prominent in the 
properties observed. Rather
than comparing absolute numbers of starburst clusters in various environments, 
it is useful to
compare star formation surface density (\Msun~{\it yr$^{-1}$}~{\it kpc$^{-2}$}) to gas 
surface mass density (\Msun~{\it pc$^{-2}$}).
In doing so, one finds that IR-selected starburst nuclei tend to exhibit a linear 
relation between
these two quantities with absolute values that imply very high star formation 
efficiences, roughly
an order of magnitude above those inferred for normal disks (see Figures 5 
and 7 in Kennicutt 1998). 
Our own GC has a very high star formation rate, given its molecular density, 
implying a high star formation efficiency. 
Given the facts described in this paper, then, we suggest that starburst clusters
tend to be produced by events which convert gas into stars at a very high 
efficiency, but the absolute
numbers of starburst clusters formed in an environment is strongly constrained 
by the available amount
of gas. In other words, our modest GC has recently formed 3 starburst clusters, 
whereas massive interacting systems, i.e. NGC 1275, have produced thousands (Carlson et al.\ 1998). 
A corollary to this suggestion is that our own
GC does not have enough molecular mass to fully populate the initial mass 
spectrum of clusters seen in {\it bona fide} starburst galaxies, where ``super-star 
clusters'' having masses ranging above 10$^5$~\Msun\ are readily observed 
(Ho \& Filippenko 1996).

\end{document}